\documentclass[letter]{aa}
\usepackage{txfonts,graphicx,aalongtable}
\usepackage{natbib}
\bibpunct{(}{)}{;}{a}{}{,} 
\usepackage{epsfig}

\newcommand{\teff}{$T_{\rm eff}$}

\begin{document}
\title{ 
HVS\,7: a chemically peculiar hyper-velocity star\thanks{Based on
observations collected at the European Southern Obser\-vatory, Paranal, Chile,
proposal 079.D-0756(A).}}
%
\author {
N. Przybilla\inst{1} \and
M. F. Nieva\inst{1,}\thanks{Current address: Max-Planck-Institut f\"ur
Astrophysik, Karl-Schwarzschild-Str. 1, D-85741 Garching, Germany} \and
A. Tillich\inst{1} \and
U. Heber\inst{1} \and
K. Butler\inst{2} \and 
W. R. Brown\inst{3}
}
\offprints{przybilla@sternwarte.uni-erlangen.de}
\institute{Dr. Remeis-Sternwarte Bamberg, Universit\"at Erlangen-N\"urnberg, Sternwartstr. 7, D-96049 Bamberg, Germany 
\and Universit\"atssternwarte M\"unchen, Scheinerstr. 1, D-81679 M\"unchen, Germany
\and Smithsonian Astrophysical Observatory, 60 Garden Street, Cambridge, MA 02138, USA
}

\date{Received... ; accepted ... }
\abstract
{Hyper-velocity stars are suggested to originate from the dynamical interaction
of binary stars with the supermassive black hole in the Galactic 
centre (GC), which accelerates one component of the binary to beyond the Galactic escape velocity.}
{The evolutionary status and GC origin of the hyper-velocity star
SDSS\,J113312.12+010824.9 (aka HVS\,7) 
is constrained from a detailed study of its stellar parameters and
chemical~composition.}
{High-resolution spectra of HVS\,7 obtained with UVES on the ESO VLT were 
analysed using state-of-the-art NLTE/LTE modelling techniques that can account for
a chemically-peculiar composition via opacity sampling.}
{Instead of the expected slight enrichments of $\alpha$-elements and near-solar iron, huge
chemical peculiarities of all elements are apparent. The helium abundance is very low 
($<$1/100 solar), C, N and O are below the detection limit, i.e they are underabundant
($<$1/100, $\lesssim$1/3 and $<$1/10 solar). Heavier elements, however, are overabundant:
the iron group by a factor of $\sim$10, P, Co and Cl by factors $\sim$40, 80 and 440 and
rare-earth elements and mercury even by $\sim$10\,000. 
An additional finding, relevant also for other chemically peculiar stars are the
large NLTE effects on abundances of \ion{Ti}{ii} and \ion{Fe}{ii} ($\sim$0.6--0.7\,dex).
The derived abundance pattern of HVS\,7 is characteristic
for the class of chemical peculiar magnetic B stars on the main sequence.
The chemical composition and high projected rotation velocity 
$v \sin i$\,$=$\,55$\pm$2\,km\,s$^{-1}$ render a 
low mass nature of HVS\,7 as a blue horizontal branch star unlikely.}
{Such a surface abundance pattern is caused by atomic diffusion in a possibly magnetically 
stabilised, non-convective atmosphere. Hence all chemical information on the star's place of 
birth and its evolution has been washed out. High precision astrometry is the only means 
to validate a GC origin for HVS\,7.}
\authorrunning {Przybilla et al.}
\maketitle
\section{Introduction}
\cite{hills88} predicted that the tidal disruption of a binary by a
supermassive black hole (SMBH) could lead to the ejection of
stars with velocities exceeding the escape velocity of our Galaxy. The Galactic
centre (GC) is the suspected place of origin of these so-called hyper-velocity
stars (HVSs), as it hosts a SMBH.
The first HVSs have only recently been discovered seren\-dipitously
\citep{brown05,hirsch05,edelmann05}.
A systematic search for such objects has resulted in the discovery of 8
additional HVSs \citep[see][]{brown07,heber08a}.

A determination of the space motion of a HVS can be used to trace back the
trajectory unambiguously to the place of ejection. This has been
done only for one HVS \citep[\object{HD\,271791}, originating from
the Galactic rim,][]{heber08a},~as sufficiently accurate proper
motions are lacking for all other HVSs at present. 

Alternatively, the chemical composition of a star may be
used to constrain its place of origin. Stars in the GC have a unique chemical 
composition, as they show enhancement of the $\alpha$-elements and roughly solar 
iron \citep[e.g.][]{cunha07}. 
So far, only one object could be studied in this way
\citep[\object{HE\,0437$-$5439}, which probably originates in the
Large Magellanic Cloud,][]{przybilla08}. This method is currently limited by 
the need to obtain high-resolution spectra of faint objects.

At present, no HVS can unambiguously be related to a GC origin. Instead,
alternative methods of acceleration such as dynami\-cal ejection from dense
cluster cores \citep{leonard91} or disruption of a binary by an
intermediate-mass black hole \citep{GPZ07} challenge the SMBH paradigm. 

\object{SDSS\,J113312.12+010824.9}\,\citep[aka HVS\,7,][]{brown06} is the 4th-brightest HVS
candidate known ($V$\,$=$\,17.80). Its spectral morphology of late B-type is consistent
with either an intermediate-mass star close to the main sequence or a
low-mass star on the blue horizontal branch \citep{brown06,heber08b}. We obtained high-resolution spectra
of HVS\,7 with UVES on the VLT and perform a quantitative spectral
analysis using state-of-the-art modelling techniques in the present work to
constrain its nature and its place of origin.

\section{Observations and quantitative analysis}
\begin{figure*}
\centering
\resizebox{.888\hsize}{!}{\includegraphics{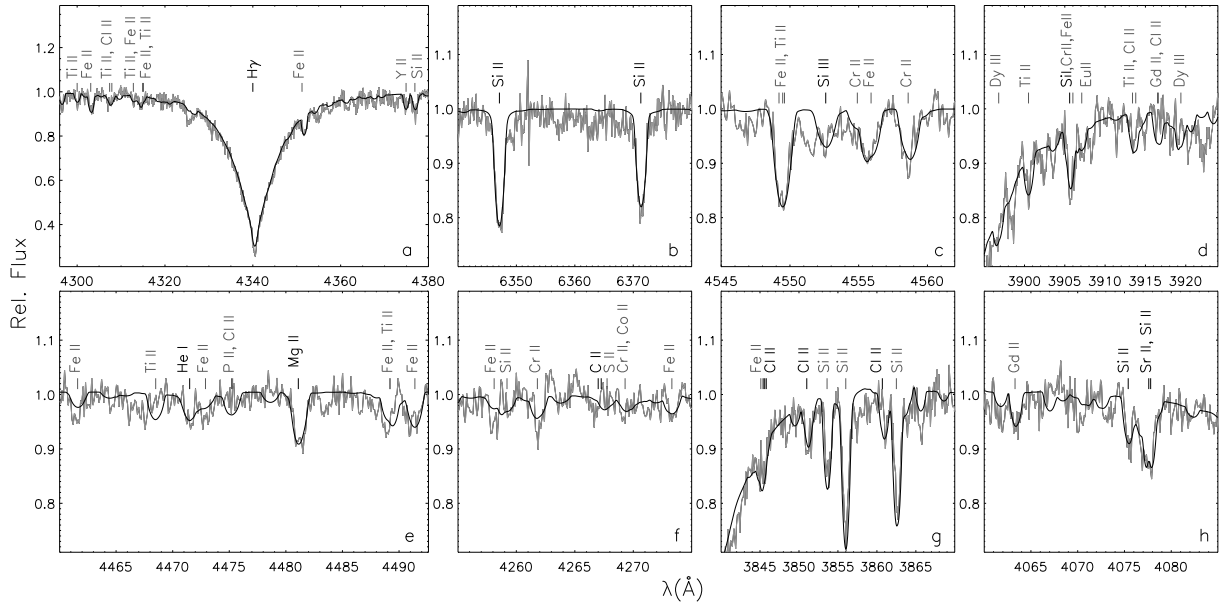}}\\[-3mm]
\caption{
Comparison of spectrum synthesis for HVS\,7 (full line, best fit for
abundances as in Table~\ref{tab_results})
with observations (grey), exemplary for some strategic regions. 
Panels (a) to (d) illustrate the {\teff} and $\log g$ determination from Balmer lines
(H$\gamma$ in (a)) and the Si ionisation equilibrium: \ion{Si}{ii} (b),
\ion{Si}{iii} (c) and \ion{Si}{i} (d). Note the strong \ion{Si}{ii} lines in
panel (b) indicative of a Si overabundance. Panel (e) illustrates the weakness 
(deficiency) of \ion{He}{i} and \ion{Mg}{ii}, (f) the absence (deficiency)
of \ion{C}{ii}, (g) the presence (strong enrichment) of \ion{Cl}{ii} and (h) a blend
of \ion{Sr}{ii} and \ion{Si}{ii}.
Iron group elements are highlighted in (c) and rare-earth elements in (d).
Other prominent spectral lines are identified.
}
\label{fits}
\end{figure*}

Seventeen spectra of HVS\,7 with a total integration time of $\sim$6.8\,h
were obtained between April 15 and May 10, 2007, covering the range of 3750 to 4950\,{\AA} and
5700 to 9000\,{\AA} at a resolving~power of $R$\,$\approx$\,35\,000.
The data reduction followed procedures described by \citet{koester01}.
A spectrum of the DC white dwarf \object{WD\,1055$-$072} was also taken so that
the continuum could be normalised reliably.
No radial velocity variations were found within the detection limits. A peak S/N of
$\approx$80 per resolution element was measured for the coadded spectrum in the blue.

An inspection of the spectrum indicated at first glance that HVS\,7 is a 
chemically-peculiar (CP) late B-type (Bp) star with moderate (projected) rotational
velocity $v \sin i$. The helium lines are extremely weak, very strong
silicon lines dominate the metal line spectrum. A closer inspection 
identified most of the $\alpha$-process elements and most of the iron group elements 
in the spectrum, and many heavier species up to rare-earth elements and mercury. 
The presence of chlorine and strong phosphorus and cobalt lines in the
spectrum of HVS\,7 is unusual even for CP stars.
Moreover, lines of carbon, nitrogen and oxygen 
are absent within the present detection limits. Many observed spectral lines
remain unidentified, in the red spectral region, in particular.
 
The quantitative analysis of HVS\,7 was carried out based on standard assumptions 
made in the field of CP stars (in particular assuming a chemically homogeneous
atmosphere). In addition, we went beyond the standard by accounting 
for NLTE effects for many important elements. We followed the
methodology discussed by \citet{przybilla06} and \citet{NiPr07}, with some
modifications and extensions in order to account for the chemical peculiarity of the
stellar atmosphere. 

In brief, NLTE line-formation
calculations were performed on the basis of line-blanketed LTE model atmospheres
\citep[{\sc Atlas12},][]{kurucz96}
using updated versions of {\sc Detail} and {\sc Surface} \citep{gid81,
butgid85}. The coupled statistical equilibrium and radiative transfer
problem was solved with {\sc Detail} and {\sc Surface} provided the formal 
solution based on the resulting NLTE occupation numbers.
Line-blocking within {\sc Detail} was realised here using an opacity
sampling technique according to \citet{kurucz96}. State-of-the-art model atoms 
were employed as indicated in Table~\ref{tab_results}, and in addition the 
hydrogen model atom of \citet{PrBu04}. 
LTE line-formation was performed with {\sc Surface}. The
list of elements for the spectrum synthesis was extended to the 
`peculiar' species for the present work. See Appendix A (available online)
for details of the line-formation calculations.

The chemical peculiarity of the stellar atmosphere required an iterative
approach to the analysis. Starting with a model with scaled solar abundances
a full analysis was performed. The resulting abundances were used to compute
an improved model with {\sc Atlas12}, which indicated corrections to stellar
parameters and abundances in a second step. Several such
iterations were required to obtain consistency and a tailored atmospheric model.

The effective temperature {\teff} and the surface gravity  $\log g$ were determined from the
Stark-broadened Balmer lines (H$\beta$ to H$_{11}$) and the ionisation equilibrium of
\ion{Si}{ii/iii} (the strongly blended \ion{Si}{i} $\lambda$3905\,{\AA}
line [in LTE] is also consistent). 
The microturbulence velocity $\xi$ was derived in the standard way by 
demanding that the line abundances within \ion{Fe}{ii} be independent of 
equivalent width. The uncertainties in the stellar parameters ({\teff}, $\log g$) were
constrained by the quality of the match of the spectral indicators within
the given S/N limitations. \cite{heber08b} determined
{\teff}\,$=$\,12\,600$\pm$500\,K, $\log g$\,$=$\,3.71$\pm$0.2 from an LTE
analysis of the Balmer series in spectra of lower resolution in
excellent agreement with the present results.

Elemental abundances were determined from visual fits to profiles of
isolated lines (or lines with resolved blends). Spectrum synthesis was mandatory 
because at the given $v \sin i$ the identification of line blends 
has to be considered carefully for a successful analysis.
Upper limits to the C, N and O abundances were determined from \ion{C}{ii}
$\lambda$4267\,{\AA}, \ion{N}{ii} $\lambda$3995\,{\AA} and \ion{O}{i}
$\lambda\lambda$7771--5\,{\AA}, the strongest predicted lines for these elements.
The only available Hg line (\ion{Hg}{ii} $\lambda$3984\,{\AA}) is
blended, therefore the Hg abundance is regarded as uncertain.

\begin{table}
\caption{Stellar parameters \& elemental abundances of HVS\,7}
\vspace{-2mm}
\label{tab_results}
\setlength{\tabcolsep}{1.46mm}
\begin{tabular}{lrrr@{\hspace{4.7mm}}lrr}
\hline
\hline
\multicolumn{2}{l}{\teff\,(K)}         & \multicolumn{2}{l}{12\,000$\pm$500} & $M/M_{\sun}$ & \multicolumn{2}{l}{3.7$\pm$0.2}\\
\multicolumn{2}{l}{$\log g$\,(cgs)}    & \multicolumn{2}{l}{3.8$\pm$0.1}     & $R/R_{\sun}$ & \multicolumn{2}{l}{4.0$\pm$0.1}\\
\multicolumn{2}{l}{$\xi$\,(km/s)}      & \multicolumn{2}{l}{3$\pm$1}         & $L/L_{\sun}$ & \multicolumn{2}{l}{300$\pm$50}\\
\multicolumn{2}{l}{$v \sin i$\,(km/s)} & \multicolumn{2}{l}{55$\pm$2} & $\tau_{\rm evol}$\,(Myr) & \multicolumn{2}{l}{150$\pm$10}\\[1mm]
\hline
\rule{0mm}{1mm}\\[-2.5mm]
Ion           & $\varepsilon^{\rm NLTE}$ & $\varepsilon^{\rm LTE}$ & \# & Ion  & $\varepsilon^{\rm LTE}$ & \#\\
\hline
\ion{He}{i}$^1$   & 8.90$\pm$0.11 & 9.17$\pm$0.11 &        2 & \ion{Ca}{ii}  & 6.60          &  1\\
\ion{C}{ii}$^2$   & $\le$6.50     & $\le$6.47     & {\ldots} & \ion{Sc}{ii}  & 4.00          &  1\\
\ion{N}{ii}$^3$   & $\le$7.50     & $\le$7.60     & {\ldots} & \ion{Cr}{ii}  & 6.40$\pm$0.12 & 10\\
\ion{O}{i}$^4$    & $\le$7.70     & $\le$8.20     & {\ldots} & \ion{Mn}{ii}  & $\le$6.50 & {\ldots}\\
\ion{Mg}{ii}$^5$  & 6.00          & 5.80          &  1 & \ion{Co}{ii}  & 6.82$\pm$0.21 &  5\\
\ion{Si}{ii}$^6$  & 8.69$\pm$0.13 & 8.61$\pm$0.15 & 12 & \ion{Sr}{ii}  & 5.00          &  1\\
\ion{Si}{iii}$^6$ & 8.65          & 8.80          &  1 & \ion{Y}{ii}   & 5.02$\pm$0.18 &  5\\
\ion{S}{ii}$^7$   & 7.35$\pm$0.12 & 7.44$\pm$0.14 &  2 & \ion{Eu}{ii}  & 5.05$\pm$0.16 &  3\\
\ion{Ti}{ii}$^8$  & 6.20$\pm$0.11 & 5.64$\pm$0.11 &  7 & \ion{Gd}{ii}  & 5.93$\pm$0.13 &  4\\
\ion{Fe}{ii}$^8$  & 8.44$\pm$0.13 & 7.78$\pm$0.16 & 26 & \ion{Dy}{ii}  & 6.00$\pm$0.23 &  5\\
\ion{P}{ii}       & {\ldots}      & 7.13$\pm$0.21 &  2 & \ion{Dy}{iii} & 4.83$\pm$0.18 &  3\\
\ion{Cl}{ii}      & {\ldots}      & 7.92$\pm$0.21 &  5 & \ion{Hg}{ii}  & 4.90         &  1\\
\hline
\end{tabular}\\
{\scriptsize
$\varepsilon(X)$\,$=$\,$\log\,(X/{\rm H})$\,+\,12. 
Error estimates consist of statistical 1$\sigma$-uncertainties
derived from line-to-line scatter (\# lines) plus 0.1\,dex for continuum
placement uncertainty (added in quadrature);
realistic uncertainties, including systematic effects e.g. if a magnetic 
field were present, are expected to be larger. 
NLTE model atoms: $^1$\cite{Pr05}; 
$^2$\cite{NiPr06,NiPr08}; $^3$\cite{PrBu01}; $^4$\cite{Pretal00}; $^5$\cite{Pretal01}; 
$^6$\cite{BeBu90}, extended \& updated; $^7$\cite{Vranckenetal96}, updated ; 
$^8$\cite{Becker98}.}
\end{table}

The results of the analysis are summarised in Table~\ref{tab_results}, the
final synthetic spectrum 
is compared to observation for some exemplary regions in Fig.~\ref{fits}.
Overall, good to excellent agreement between the model and observation is achieved.
Note that many spectral lines are strongly blended and were omitted from the 
analysis though a good match is obtained in the
comparison with the complete synthetic spectrum, with a few exceptions. 
In addition, several observed lines are unaccounted for by the model.
No positive identification of these lines could be found,
a common problem in the field of CP stars, because the atomic data are
unavailable. These transitions supposedly arise from highly-excited levels of iron
group species, which are unobserved at normal abundance values, or are
lines of heavier elements.

\begin{figure}
\centering
\resizebox{.88\hsize}{!}{\includegraphics{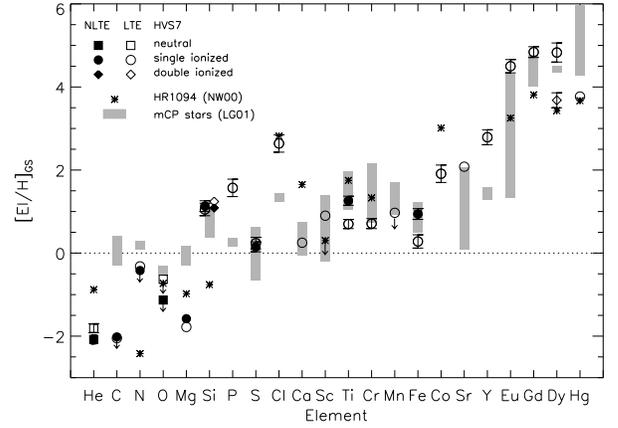}}
\caption{Abundance pattern of HVS\,7 
\citep[relative to solar values,][]{GrSa98} in comparison to
magnetic CP stars. Symbols are identified in the legend.
The error bars represent 
values from Table~\ref{tab_results}. Realistic errors, accounting for
systematic effects (e.g. if a magnetic field were present), are expected 
to be larger. Upper limits are indicated by arrows.
Asterisks mark abundances for the Cl-Co-rich B9p star
HR\,1094 \citep{NiWa00} and grey bars abundance ranges observed in a sample of magnetic CP
stars \citep{LoGaetal01}.
\label{abundances}}
\end{figure}

Many chemical peculiarities relative to the solar
composition \citep{GrSa98} are found, as shown in Fig.~\ref{abundances}. In general,
we find an increase in abundance with atomic weight. The helium abundance is extremely 
low ($<$1/100 solar), C,
N and O are absent within the detection limits ($<$1/100, $\lesssim$1/3,
$<$1/10 solar, respectively) and magnesium is markedly underabundant ($\sim$1/30 solar).
Silicon is the most abundant metal ($>$10$\times$solar) and chlorine, cobalt
and phosphorus are
pronouncedly overabundant ($\sim$440, 80 and 40$\times$solar).
All heavier elements are largely overabundant: $\sim$10$\times$solar
values for iron group elements and $\sim$10\,000$\times$solar values for
rare-earth elements and mercury.

\paragraph{NLTE effects.}
There are very few NLTE analyses of CP stars available in the literature,
which mostly deal with $\lambda$\,Boo stars \citep{paunzen99,kamp01}.
We therefore want to briefly discuss some results that may be of relevance to other 
studies of Bp objects and possibly for other CP star groups.

NLTE and LTE abundances from the weak \ion{He}{i} lines (only triplet lines are
observed) show differences up to 0.3\,dex (\ion{He}{i} $\lambda$4471\,{\AA}), 
as a significant NLTE overpopulation in the triplet states can build up deep 
in the atmosphere because radiative de-excitation to the ground state is forbidden.
The same mechanism is in operation for \ion{O}{i} $\lambda\lambda$7771--5\,{\AA}.
Similar cases may be
\ion{P}{ii} and \ion{Cl}{ii} (where we lack model atoms for NLTE
calculations). Lines from spin systems which are not connected radiatively to
the ground state are observed for both ions, while lines from the spin
systems that include the ground state remain below the detection limit, 
although both should be visible when assuming LTE.
NLTE effects for \ion{C}{ii} $\lambda$4267\,{\AA}, \ion{N}{ii}
$\lambda$3995\,{\AA} (for upper limit abundances), \ion{Mg}{ii} and \ion{S}{ii} remain
small, as the lines are formed deep in the atmosphere where level
populations are close to detailed equilibrium. The strongest metal lines in the spectrum, of
\ion{Si}{ii}, are often saturated and  the NLTE effects remain small
because the lines become optically thick.

Differences between the NLTE and LTE analyses are mostly similar to those in
main sequence stars of normal composition \citep{hempel03}, except for \ion{Ti}{ii} 
and \ion{Fe}{ii}. For these, surprisingly large NLTE effects are
found here, $\sim$0.6--0.7\,dex, exceeding even the effects found in 
supergiants \citep{przybilla06} at similar {\teff}.
The main NLTE mechanisms in these cases are an overionisation of the ground state and
the energetically low levels. Because of the large overabundances the lines
form at much smaller depths in the atmosphere than under
normal conditions (though avoiding
saturation), so that the level populations deviate strongly from detailed
equilibrium values.
Such a mechanism is also plausible for
the first ions of the rare-earth elements, which have ionisation energies
of about 10-12\,eV, i.e. they are efficiently ionised by the intense
radiation field longward of the Ly jump. This may e.g. explain the virtual
absence of expected \ion{Gd}{ii} resonance lines in the optical, whereas
lines from excited \ion{Gd}{ii} levels are observed. Accounting for NLTE effects may also
help to establish the \ion{Dy}{ii/iii} ionisation~equilibrium. 

For the moment, we can conclude that some CP stars with
elemental overabundances, such as the
Bp star analysed here, may be good candidates for the study of NLTE effects. Further
investigations will be required, as the present results cannot be
generalised to other CP star classes.

\section{Discussion}

\begin{figure}
\centering
\resizebox{.785\hsize}{!}{\includegraphics{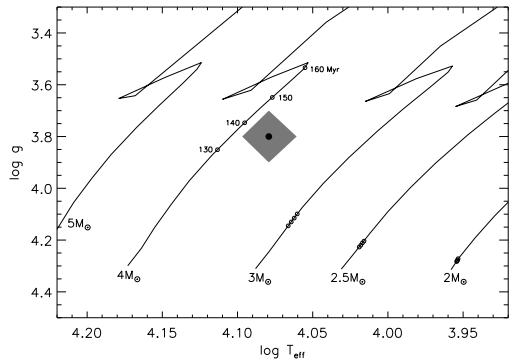}}\\[-3mm]
\caption{Position of HVS\,7 in the $T_{\rm eff}$--$\log g$ diagram for the
determination of its evolutionary mass and age (error box in grey).
Evolutionary tracks for solar metallicity are adopted from \citet{schaller92}. 
Time markers in the range 130--160\,Myr are shown for each track. }
\label{tracks}
\end{figure}

Based on the position of HVS\,7 in a {\teff}--$\log g$-diagram alone one cannot
decide whether the star is an intermediate-mass star of 3--4\,$M_{\sun}$ 
close to the main sequence or a low-mass star of $\sim$0.5--0.7\,$M_{\sun}$
on the horizontal branch. \cite{heber08b} show the problem in their Fig.~2.
 An important constraint is
also the rotational velocity, which is in general low ($<$8\,km\,s$^{-1}$) for hot horizontal
branch stars \citep[\teff$\gtrsim$11\,000\,K,][]{behr03}. The high
$v \sin i$ of 55\,km\,s$^{-1}$ does not necessarily rule out an HB star, as the star
might have been spun up by tidal locking to the orbital motion before the binary
was disrupted \citep[see][]{hansen07} to release the HVS.
Chemical peculiarities can occur both on the main
sequence \citep[e.g.][]{smith96} and on the horizontal branch
\citep[e.g.][]{moehler01,behr03}, though the density of metal lines in the
latter case is much lower than observed for HVS\,7. 
Hence, HVS\,7 is very likely a main-sequence~star.

A comparison with evolutionary tracks \citep{schaller92} 
allows the mass of HVS\,7 to be constrained 
(Fig.~\ref{tracks}), 
and other fundamental stellar parameters such as radius
$R$ and luminosity $L$, and its evolutionary lifetime $\tau_{\rm evol}$ 
to be determined, see Table~\ref{tab_results}. The spectroscopic distance is then 
59$\pm$6\,kpc, in excellent agreement with the distance estimate of \cite{brown06}.
Note that the chemical peculiarities affect only the outer layers of the star. Moreover, 
they have developed with time, so that evolution tracks for standard composition are appropriate.

An identification of HVS\,7 as a CP star comes not entirely unexpected in
view of the commonness of late B-type stars with abundance peculiarities.
In the following we attempt to classify HVS\,7 within the zoo of CP stars. The
dominance of the Si lines defines HVS\,7 as a hot Bp star \citep[e.g.][]{smith96}, indicating a
connection to magnetic stars. Indeed, the observed chemical peculiarities for the elements 
heavier than chlorine resemble in most cases those usually found for magnetic CP stars \citep[e.g. the sample
discussed by][]{LoGaetal01}, see Fig.~\ref{abundances}. However, the lighter elements
deviate in many respect from this behaviour. We therefore include also
\object{HR\,1094} in the
comparison, a rare member of the Cl-Co-rich stars \citep{NiWa00} with similar
stellar parameters as HVS\,7. Noteworthy is also the He-weakness. 
We conclude that HVS\,7 is an exceptional member of the Bp stars, showing
characteristics of several other subclasses of CP stars. The abundance
peculiarities may be understood in the standard framework for the
theoretical explanation of CP stars: via atomic diffusion processes, i.e.
the interplay of gravitational settling and radiative levitation, 
in magnetically stabilised, non-convective atmospheres \citep[e.g.][]{michaud70}. 
 
As HVS\,7 might be a magnetic star, we can constrain its field strength 
from extra broadening of metal lines (in addition to rotation) due to 
the Zeeman effect. Assuming that this broadening can be mimicked by 
microturbulence, we derive an upper limit of 3\,kG using Eqns. 1 \& 2 of  \cite{kupka96}. 
This may put valuable constraints on
magnetic fields in the GC if HVS\,7 could be proven to be of GC origin (see
below) and the field as being of fossil origin \citep[note that the
star S2 near the GC is also suggested to be magnetic,][]{martins08}.
A neglect of Zeeman splitting results in the overestimation of
abundances, an effect most pronounced for the rare-earth elements. 
In addition, chemical surface inhomogeneities  (e.g. spots) may be caused 
by a magnetic field. In this case the abundances 
derived here would be averaged surface abundances.\\
The aim of our spectral analysis was to search for chemical signatures that
might constrain the place of birth of HVS\,7 in the Galaxy, e.g. in the GC.
However, diffusion processes have erased any indication of its original composition. 

A verification of the Hills mechanism for the acceleration of HVS\,7 has 
to await precise astrometry. Proper motion measurements have to be 
combined with the known radial velocity to obtain its space velocity 
vector and to reconstruct the trajectory of HVS\,7 back to its place of 
birth. Assuming that HVS\,7 originated in the GC, we can calculate the 
corresponding proper motion components as described by \citet{edelmann05},
resulting in $\mu_{\alpha}$\,$=$\,$-$0.53\,mas\,yr$^{-1}$,
$\mu_{\delta}$\,$=$\,$-$0.51\,mas\,yr$^{-1}$ and a space velocity of
$\sim$830\,km\,s$^{-1}$. The corresponding time of flight (120\,Myr) from the
GC to its present location is in accordance with the evolutionary life time of 150$\pm$10\,Myr. 
However, the predicted proper motion is so small that present-day
instrumentation is not capable of measuring it with sufficient precision.
Final conclusions can be drawn only after the GAIA mission.

\begin{acknowledgements}
We thank the staff of the ESO Paranal observatory for their support
with the observations, M.~Firnstein for help with the data reduction and
F.~Kupka for a valuable discussion on magnetic fields in CP stars.
M.~F.~N. and A.~T. gratefully acknowledge financial support by the Deutsche
Forschungsgemeinschaft (grant HE\,1356/45-1).
\end{acknowledgements}

\Online

\begin{appendix}
\section{Spectral line analysis}
In this appendix we provide details on our spectral line analysis of HVS\,7.
Table~\ref{taba1} summarises our line data and the results from the
abundance analysis of individual lines. The first columns give the  
wavelength $\lambda$ (in {\AA}), excitation energy of the lower level $\chi$ (in eV), adopted
oscillator strength $\log gf$, an accuracy flag for the oscillator strength,
the source of the $\log gf$ value and the reference for Stark broadening
parameters (tabulations by Sch\"oning \& Butler based on the theory of Vidal et al.~1973
were used for the Balmer lines). Then, the derived abundances
$\varepsilon$\,$=$\,$\log (X/{\rm H})$\,$+$\,12 in NLTE and LTE are tabulated. 

Radiative damping
constants were preferentially calculated from level lifetimes, adopted from the Opacity
Project (Seaton et al.~1994) and other sources in the (astro-)physics
literature, where available. In other cases data were adopted from Kurucz \& Bell (1995) 
or calculated assuming classical damping. Note that radiative damping is
negligible in the present case, such that we omit a more detailed discussion. Van-der-Waals broadening
was neglected, as the atmospheric plasma is well ionised throughout the
line-formation region. Resonance broadening for hydrogen was also neglected
for the same reason.

Finally, note that isotopic and hyperfine-structure for \ion{Eu}{ii} and
\ion{Hg}{ii} lines were treated according to Lawler et al. (2001) and
Castelli \& Hubrig (2004), respectively.\\[1cm]
{\large\sffamily\bfseries Online References}\\[3mm]
{\tiny
\sloppy\clubpenalty4000\widowpenalty4000%
Barnard, A. J., Cooper, L., \& Shamey, L. J.~1969, \aap, 1, 28\\
Bates, D., \& Damgaard, A.~1949, Phil. Trans. R. Soc. London, Ser. A, 242, 101\\
Bi\'emont, E., \& Lowe, R. M. 1993, \aap, 273, 665\\
Castelli, F. \& Hubrig, S. 2004, \aap, 425, 263\\
Cowley, C.~1971, Observatory, 91, 139\\
Den Hartog, E. A., Lawler, J. E., Sneden, C., \& Cowan, J. J. 2006, \apjs, 167,\\ \rule{3mm}{0mm}292\\
Fernley, J. A., Taylor, K. T., \& Seaton, M. J. 1987, J. Phys. B, 20, 6457\\
Froese Fischer, C., Tachiev, G., \& Irimia, A. 2006, At. Data Nucl. Data Tables,\\ \rule{3mm}{0mm}92, 607\\
Fuhr, J. R., \& Wiese, W. L.~1998, in CRC Handbook of Chemistry and Physics,\\ \rule{3mm}{0mm}79th ed., ed. D. R. Lide (CRC Press, Boca Raton)\\
Fuhr, J. R., Martin, G. A., \& Wiese, W. L.~1988, J. Phys. \& Chem. Ref. Data,\\ \rule{3mm}{0mm}Vol.17, Suppl. 4\\
Griem, H. R.~1964, Plasma Spectroscopy (McGraw-Hill Book Com\-pany, New\\ \rule{3mm}{0mm}York)\\
Griem, H. R.~ 1974, Spectral Line Broadening by Plasmas (Academic Press,\\ \rule{3mm}{0mm}New York and London)\\
Hannaford, P., Lowe, R. M., Grevesse, N., Bi\'emont, E., \& Whaling, W. 1982,\\ \rule{3mm}{0mm}\apj, 261, 736\\
Kurucz,\,R.\,L.\,1988,\,in\,Trans.\,IAU\,XXB,\,ed.\,M.\,McNally\,(Kluwer,Dordrecht),168\\
Kurucz, R.\,L. 1994a, Kurucz CD-ROM 20 (SAO, Cambridge, Mass.)\\
Kurucz, R.\,L. 1994b, Kurucz CD-ROM 22 (SAO, Cambridge, Mass.)\\
Kurucz, R.\,L., \& Bell, B.~1995, Kurucz CD-ROM 23 (SAO, Cambridge, Mass.)\\
Lanz, T., Dimitrijevi\'c, M. S., \& Artru, M.-C.~1988, \aap, 192, 249\\
Lawler, J. E., Wickliffe, M. E., den Hartog, E. A., \& Sneden, C. 2001, \apj, 563,\\ \rule{3mm}{0mm}1075\\
Martin, G. A., Fuhr, J. R., \& Wiese, W. L.~1988, J. Phys. \& Chem. Ref. Data,\\ \rule{3mm}{0mm}Vol.17, Suppl. 3\\
Matheron, P., Escarguel, A., Redon, R., Lesage, A., \& Richou, J. 2001, J. Quant.\\ \rule{3mm}{0mm}Spectrosc. Radiat. Transfer, 69, 535\\
Mendoza, C., Eissner, W., Le Dourneuf, M., \& Zeippen, C. J. 1995, J. Phys. B,\\ \rule{3mm}{0mm}28, 3485\\
Nahar, S. N. 1998, At. Data Nucl. Data Tables, 68, 183\\
Seaton, M. J., Yan, Y., Mihalas, D., Pradhan, A. K.~1994, MNRAS, 266, 805\\
Shamey, L. J. 1969, Ph. D. Thesis, University of Colorado\\
Vidal, C. R., Cooper, J., \& Smith, E. W.~1973, \apjs, 25, 37\\
Wiese, W. L., Smith, M. W., \& Miles, B. M.~1969, Nat. Stand. Ref. Data Ser.,\\ \rule{3mm}{0mm}Nat. Bur. Stand. (U.S.), NSRDS-NBS 22, Vol. II\\
Wiese, W. L., Fuhr, J. R., \& Deters, T. M.~1996, J. Phys. \& Chem. Ref. Data,\\ \rule{3mm}{0mm}Mon. 7\\
Younger, S. M., Fuhr, J. R., Martin, G. A., \&  Wiese, W. L. 1978, J. Phys.\&\\ \rule{3mm}{0mm}Chem. Ref. Data, 7, 495\\
Zhang, Z. G., Svanberg, S., Palmeri, P., Quinet, P., \& Bi\'emont, E. 2002,\mnras,\\ \rule{3mm}{0mm}334, 1\\
}

{\footnotesize
\begin{longtable}{lrrlrrrr}
\caption{Spectral line analysis\label{taba1}}\\
\hline\hline
$\lambda$\,({\AA}) & $\chi$\,(eV) & $\log gf$ & Acc. & Src. & Broad. & $\varepsilon^{\rm NLTE}$ & $\varepsilon^{\rm LTE}$\\
\hline
\endfirsthead
\caption{continued.}\\
\hline\hline
$\lambda$\,({\AA}) & $\chi$\,(eV) & $\log gf$ & Acc. & Src. & Broad. & $\varepsilon^{\rm NLTE}$ & $\varepsilon^{\rm LTE}$\\
\hline
\endhead
\hline
\endfoot
\ion{He}{i}:\\
4026.18 &   20.96 & $-$2.60 & A  & FTS & S   &       8.90 &       9.13 \\
4026.19 &   20.96 & $-$0.63 & A  & FTS & S\\
4026.20 &   20.96 & $-$0.85 & A  & FTS & S\\
4026.36 &   20.96 & $-$1.33 & A  & FTS & S\\
4471.47 &   20.96 & $-$0.21 & A  & FTS & BCS &       8.90 &       9.20 \\
4471.49 &   20.96 & $-$0.43 & A  & FTS & BCS\\
4471.68 &   20.96 & $-$0.91 & A  & FTS & BCS\\[2mm]
\ion{C}{ii}:\\
4267.000 &  18.04 &    0.56 & C+ & WFD & G   & $\leq$6.50 & $\leq$6.47 \\
4267.260 &  18.04 &    0.74 & C+ & WFD & G\\[2mm]
\ion{N}{ii}:\\
3995.00  &  18.50 &    0.21 & B  & WFD & C   & $\leq$7.50 & $\leq$7.60 \\[2mm]
\ion{O}{i}:\\
7771.94  &   9.15 &    0.37 & A  & WFD & C   & $\leq$7.70 & $\leq$8.20 \\
7774.17  &   9.15 &    0.22 & A  & WFD & C\\
7775.39  &   9.15 &    0.00 & A  & WFD & C\\[2mm]
\ion{Mg}{ii}:\\
4481.126 &   8.86 &    0.73 & B  & FW  & G   &       6.00 &       5.80 \\
4481.150 &   8.86 & $-$0.57 & B  & FW  & G\\
4481.325 &   8.86 &    0.57 & B  & FW  & G\\[2mm]
\ion{Si}{ii}:\\
3853.665 &   6.86 & $-$1.40 & C+ & FFTI& LDA &       8.50 &       8.48\\
3856.018 &   6.86 & $-$0.45 & C+ & FFTI& LDA &       8.60 &       8.60\\
3862.595 &   6.86 & $-$0.71 & C+ & FFTI& LDA &       8.60 &       8.60\\
3954.300 &  12.53 & $-$1.11 & E  & N98 & C   &       8.75 &       8.63\\
3954.504 &  12.53 & $-$0.94 & E  & N98 & C\\
4075.452 &   9.84 & $-$1.40 & C+ & MER & C   &       8.70 &       8.48\\
4076.780 &   9.84 & $-$1.70 & C+ & MER & C\\
4200.658 &  12.53 & $-$0.89 & E  & N98 & C   &       8.70 &       8.54\\
4200.887 &  12.53 & $-$2.03 & E  & N98 & C\\
4200.898 &  12.53 & $-$0.73 & D  & N98 & C\\
4376.969 &  12.84 & $-$0.89 & E  & N98 & C   &       8.80 &       8.66\\
4376.994 &  12.84 & $-$1.02 & E  & N98 & C\\
4621.418 &  12.53 & $-$0.61 & D  & MELZ& C   &       8.80 &       8.59\\
4621.696 &  12.53 & $-$1.75 & E  & MELZ& C\\
4621.722 &  12.53 & $-$0.45 & D  & MELZ& C\\
5957.561 &  10.07 & $-$0.35 & D  & WSM & LDA &       8.70 &       8.52\\
5978.929 &  10.07 & $-$0.06 & D  & WSM & LDA &       8.70 &       8.56\\
6347.103 &   8.12 &    0.17 & C+ & FFTI& LDA &       8.70 &       8.82\\
6371.359 &   8.12 & $-$0.13 & C+ & FFTI& LDA &       8.75 &       8.80\\[2mm]
\ion{Si}{iii}:\\
4552.622 &  19.02 &    0.29 & B+ & FFTI& C   &       8.65 &       8.80\\[2mm] 
\ion{P}{ii}:\\
4420.712 &  11.02 & $-$0.33 & D  & CA  & C   &     \ldots &       7.00\\
4499.230 &  13.38 &    0.47 & D  & CA  & C   &     \ldots &       7.25\\[2mm]
\ion{S}{ii}:\\
4524.675 &  15.07 & $-$0.94 & D  & FW  & C   &       7.30 &       7.37\\
4524.941 &  15.07 &    0.17 & D  & FW  & C\\
4815.552 &  13.67 &    0.09 & D  & FW  & C   &       7.40 &       7.50\\[2mm]
\ion{Cl}{ii}:\\
3845.362 &  15.95 &    0.02 & D  & WSM & C   &     \ldots &     8.00\\
3845.639 &  15.95 &    0.12 & D  & WSM & C\\
3845.788 &  15.95 & $-$0.22 & D  & WSM & C\\
3850.988 &  15.95 &    0.45 & D  & WSM & C   &     \ldots &     8.00\\
3851.374 &  15.95 &    0.25 & D  & WSM & C\\
3851.651 &  15.95 & $-$0.35 & D  & WSM & C\\
3860.828 &  15.96 &    0.74 & D  & WSM & C   &     \ldots &     7.60\\
3860.990 &  15.96 &    0.14 & D  & WSM & C\\
3861.378 &  15.96 & $-$0.70 & D  & WSM & C\\
4794.556 &  13.38 &    0.40 & C  & FW  & C   &     \ldots &     7.90\\
4810.070 &  13.38 &    0.24 & C  & FW  & C   &     \ldots &     8.10\\
\clearpage
\ion{Ca}{ii}:\\
3933.663 &   0.00 &    0.14 & C  & FW  & G   &     \ldots &     6.60\\[2mm]
\ion{Sc}{ii}:\\
4246.822 &   0.32 &    0.28 & D  & MFW & C   &     \ldots &     4.00\\[2mm]
\ion{Ti}{ii}:\\
3900.551 &   1.13 & $-$0.45 & D  & MFW & C   &       6.20 &     5.61\\
3913.468 &   1.12 & $-$0.53 & D  & MFW & C   &       6.10 &     5.55\\
4300.049 &   1.18 & $-$0.77 & D$-$& MFW& C   &       6.20 &     5.63\\
4301.914 &   1.16 & $-$1.16 & D$-$& MFW& C   &       6.20 &     5.67\\
4395.033 &   1.08 & $-$0.66 & D$-$& MFW& C   &       6.20 &     5.65\\
4443.794 &   1.08 & $-$0.70 & D$-$& MFW& C   &       6.20 &     5.65\\
4911.193 &   3.12 & $-$0.34 & D  & MFW & C   &       6.30 &     5.75\\[2mm]
\ion{Cr}{ii}:\\
4110.990 &   3.76 & $-$2.02 & X  & K88 & C   &     \ldots &     6.50\\
4111.003 &   3.10 & $-$1.92 & X  & K88 & C\\
4242.364 &   3.87 & $-$1.33 & X  & YFMW& C   &     \ldots &     6.50\\
4275.567 &   3.86 & $-$1.71 & X  & K88 & C   &     \ldots &     6.30\\
4558.650 &   4.07 & $-$0.66 & D  & MFW & C   &     \ldots &     6.40\\
4588.199 &   4.07 & $-$0.63 & D  & MFW & C   &     \ldots &     6.40\\
4616.629 &   4.07 & $-$1.29 & D  & MFW & C   &     \ldots &     6.40\\
4618.803 &   4.07 & $-$1.11 & D  & MFW & C   &     \ldots &     6.40\\
4634.070 &   4.07 & $-$1.24 & D  & MFW & C   &     \ldots &     6.40\\
4824.127 &   3.87 & $-$0.96 & X  & K94a& C   &     \ldots &     6.30\\
4876.399 &   3.86 & $-$1.47 & D  & MFW & C   &     \ldots &     6.35\\[2mm]
\ion{Mn}{ii}:\\
4755.727 &   5.40 & $-$1.24 & X  & K88 & C   &     \ldots &$\leq$6.50\\[2mm]
\ion{Fe}{ii}:\\
3863.985 &   4.15 & $-$3.43 & X  & K88 & C   &       8.50 &     7.95\\
3872.766 &   2.70 & $-$3.32 & X  & K88 & C   &       8.40 &     7.80\\
3935.962 &   5.57 & $-$1.86 & D  & FMW & C   &       8.50 &     7.92\\
4004.080 &  10.68 & $-$3.72 & X  & K88 & C   &       8.40 &     7.80\\
4173.461 &   2.58 & $-$2.18 & C  & FMW & C   &       8.45 &     7.65\\
4178.862 &   2.58 & $-$2.47 & C  & FMW & C   &       8.40 &     7.65\\
4273.326 &   2.70 & $-$3.34 & D  & FMW & C   &       8.50 &     7.90\\
4303.176 &   2.70 & $-$2.49 & C  & FMW & C   &       8.45 &     7.65\\
4351.768 &   2.70 & $-$2.10 & C  & FMW & C   &       8.20 &     7.40\\
4385.387 &   2.78 & $-$2.57 & D  & FMW & C   &       8.40 &     7.70\\
4489.183 &   2.83 & $-$2.97 & D  & FMW & C   &       8.50 &     7.85\\
4491.405 &   2.86 & $-$2.69 & C  & FMW & C   &       8.40 &     7.70\\
4508.288 &   2.86 & $-$2.30 & X  & K94b& C   &       8.40 &     7.70\\
4520.224 &   2.81 & $-$2.61 & D  & FMW & C   &       8.45 &     7.70\\
4522.634 &   2.84 & $-$2.11 & X  & K94b& C   &       8.50 &     7.75\\
4549.474 &   2.83 & $-$1.75 & C  & FMW & C   &       8.60 &     7.85\\
4555.893 &   2.83 & $-$2.33 & X  & K94b& C   &       8.40 &     7.65\\
4576.340 &   2.84 & $-$3.04 & D  & FMW & C   &       8.40 &     7.77\\
4629.339 &   2.81 & $-$2.38 & D  & FMW & C   &       8.30 &     7.60\\
4635.316 &   5.96 & $-$1.65 & D  & FMW & C   &       8.40 &     7.83\\
4666.758 &   2.83 & $-$3.34 & D  & FMW & C   &       8.60 &     8.00\\
4731.453 &   2.89 & $-$3.37 & D  & FMW & C   &       8.50 &     7.90\\
4923.927 &   2.89 & $-$1.32 & C  & FMW & C   &       8.45 &     7.80\\
6317.983 &   7.47 & $-$1.99 & X  & K88 & C   &       8.35 &     7.80\\
6516.081 &   2.89 & $-$3.45 & D  & FMW & C   &       8.50 &     7.90\\
6518.774 &  12.89 &    0.35 & X  & K88 & C   &       8.40 &     7.95\\[2mm]
\ion{Co}{ii}:\\
4160.673 &   3.41 & $-$1.83 & X  & K88 & C   &   \ldots   &     6.50\\
4497.431 &   3.41 & $-$2.54 & X  & K88 & C   &   \ldots   &     6.90\\
4516.633 &   3.46 & $-$2.56 & X  & K88 & C   &   \ldots   &     6.90\\
4660.656 &   3.36 & $-$2.21 & X  & K88 & C   &   \ldots   &     6.90\\
4831.242 &   5.05 & $-$1.67 & X  & K88 & C   &   \ldots   &     6.90\\[2mm]
\clearpage
\ion{Sr}{ii}:\\
4215.520 &   0.00 & $-$0.17 & X  & FW  & C   &   \ldots   &     5.00\\[2mm]
\ion{Y}{ii}:\\
3950.356 &   0.10 & $-$0.49 & X  & HLG & C    &  \ldots   &     5.00\\
4177.636 &   0.41 & $-$0.16 & X  & HLG & C    &  \ldots   &     5.10\\
4374.946 &   0.41 &    0.16 & X  & HLG & C    &  \ldots   &     4.80\\
4398.010 &   0.13 & $-$1.00 & X  & HLG & C    &  \ldots   &     5.00\\
4900.110 &   1.03 & $-$0.09 & X  & HLG & C    &  \ldots   &     5.20\\[2mm]
\ion{Eu}{ii}:\\
3819.67  &   0.00 &    0.51 & X  & LWDS& C    &  \ldots   &      4.90\\
3907.11  &   0.21 &    0.17 & X  & LWDS& C    &  \ldots   &      5.10\\
4205.04  &   0.00 &    0.21 & X  & LWDS& C    &  \ldots   &      5.15\\[2mm]
\ion{Gd}{ii}:\\
3916.509 &   0.60 & $-$0.04 & X  & DLSC& C    &  \ldots   &      5.80\\
4037.323 &   0.66 & $-$0.11 & X  & DLSC& C    &  \ldots   &      5.95\\
4037.893 &   0.56 & $-$0.42 & X  & DLSC& C\\
4063.384 &   0.99 &    0.33 & X  & DLSC& C    &  \ldots   &      5.95\\
4184.258 &   0.49 &    0.00 & X  & DLSC& C    &  \ldots   &      6.00\\[2mm]
\ion{Dy}{ii}:\\
3872.11  &   0.00 &    0.00 & X  & BL  & C    &  \ldots   &      5.70\\
3944.68  &   0.00 &    0.10 & X  & BL  & C    &  \ldots   &      6.00\\
3996.69  &   0.59 & $-$0.20 & X  & BL  & C    &  \ldots   &      6.00\\
4000.45  &   0.10 &    0.06 & X  & BL  & C    &  \ldots   &      6.00\\
4050.56  &   0.59 & $-$0.42 & X  & BL  & C    &  \ldots   &      6.30\\[2mm]
\ion{Dy}{iii}:\\
3896.817 &   0.52 & $-$1.10 & X  & ZSP & C    &  \ldots   &      4.80\\
3919.398 &   0.52 & $-$1.34 & X  & ZSP & C    &  \ldots   &      4.70\\
3930.632 &   0.00 & $-$0.88 & X  & ZSP & C    &  \ldots   &      5.00\\[2mm]
\ion{Hg}{ii}:\\
3983.93  &   4.40 & $-$1.52 & X  & CH  & C    &  \ldots   &      4.90
\end{longtable}
}
{\footnotesize\noindent
accuracy indicators -- uncertainties within: A: 3\%; B: 10\%; C:
25\%; D: 50\%; E: larger than 50\%; X: unknown\\[1.5mm]
sources of $gf$-values -- 
BL: Bi\'emont \& Lowe (1993); 
CA: Coulomb approximation (Bates \& Damgaard 1949);
CH: Castelli \& Hubrig (2004);
DLSC: Den Hartog et al. (2006);
FFTI: Froese Fischer et al. (2006); 
FTS: Fernley et al. (1987); 
FMW: Fuhr et al. (1988); 
FW: Fuhr \& Wiese (1998); 
HLG: Hannaford et al. (1982);
K88: Kurucz (1988);
K94a: Kurucz (1994a);
K94b: Kurucz (1994b);
LWDS: Lawler et al. (2001);
MELZ: Mendoza et al. (1995); 
MER: Matheron et al. (2001); 
MFW: Martin et al. (1988); 
N98: Nahar (1998); 
WFD: Wiese et al. (1996); 
WSM: Wiese et al. (1969), when available replaced by improved $gf$-values from Fuhr \& Wiese (1998);
YFMW: Younger et al. (1978);
ZSP: Zhang et al. (2002)\\[1.5mm]
sources for Stark broadening parameters -- BCS: Barnard et al.~(1969); C:
Cowley (1971); G: Griem~(1964, 1974); LDA: Lanz et al. (1988); S: Shamey
(1969)
}

\end{appendix}

\end{document}